============================Cut Here===========================================
\documentstyle[aps]{revtex}
\title
{Effective two-body interactions in the $ s-d$ shell nuclei from sum
rules equations in tranfer reactions}
\author
{H.Sharda\thanks{shardah@usa.net}, Bansal R. K. }
\address
{{\it Department of Physics, Panjab University,}\\
{\it Chandigarh -160014, India}\\[12pt]
{\it and} \\
Kumar Ashwani\\
{\it  Department of Applied Sciences, Punjab Engineering
College,}\\
{\it Chandigarh - 160012, India}}

\begin{document}

\maketitle
\tightenlines

\begin{abstract}
Average effective two-body interaction matrix elements in the
s-d shell have been extracted, from data on experimentally
measured isospin centroids, by combining the recently derived
new sum rules  equations for pick-up reactions with similar known equations
for stripping reactions performed on general multishell target
states. Using this combination of stripping and pick-up equations, the
average effective matrix elements for the shells, $1d^{2}_{5/2}$, 
$2s^{2}_{1/2}$ and $1d^{2}_{3/2}$ respectively have been obtained. A new 
feature of the present work is that the restriction imposed in earlier
works on target states, that it be populated only by active neutrons has
now been abandoned.
\end{abstract}
\vfill \eject

\narrowtext
\section*{I. INTRODUCTION}
Effective   nucleon-nucleon    interaction    forms    an
important information input in  any  shell-model  calculation  of
nuclear energy spectrum. However, the exact form of nuclear force
being unknown, it is in itself quite a  problem  to  obtain  the 
matrix elements of effective   two-body  interaction.  A  common 
approach is to parametrise the interaction and then  obtain  the
parameters by making fits to  experimental  data.   Detailed 
calculations  of  energy spectra are required for this purpose
which may involve the setting  up  and diagonalisation of large
matrices.\\
An alternative approach is  to  obtain  meaningful  averages  of 
effective interaction matrix elements in  a  simpler  manner  by 
making use of the  energy-weighted sum rules [1,2,3].
These sum rules relate the energy centroids of residual  nucleus 
states populated via a direct single particle transfer reaction to the average
effective  two-body interaction matrix elements. So far, sum rules for
stripping reactions have been applied more widely
than those for pick-up reactions. The complexity  of  the  expression  for
D
centroids, derived from monopole energy-weighted sum rule,
increases as we move from the simplest case of inequivalent 
transfer [2], to the case of equivalent transfer. In a
situation when  the  target  state  contains  only neutrons in
the transfer orbit, and an addition of a single particle to it is  being
studied,  the expressions for isospin
centroids are still fairly simple [3] and have been applied to a 
large number of cases [4,5]. When there is no restriction on the
occupancy of the transfer orbit in the  target  state,  the
monopole  sum  rule  for  the stripping case [6] assumes
more   complexity    by   way   of   a   two-body    correlation 
term,$<H^{01}>$, which defies analytical
evaluation.  For meaningful application of the sum rules, it  is 
essential to eliminate this term in an appropriate manner.\\
Recently, explicit algebraic expressions have been reported [7]  for
isospin centroids
of  residual  nucleus  states  obtained   via   single   nucleon 
equivalent pick-up reactions
from  general multishell target states. These along with earlier
similar expressions for centroids in the stripping case form a
set which has been used [7] to obtain the $1f^{2}_{7/2}$ matrix
elements after eliminating the
`problem term' $<H^{01}>$.\\
The present article presents the results of a similar approach
extending the application of these sum rules to stripping and
pick-up reactions in the s-d shell nuclei as targets.  For the
sake of completeness, the expressions for isospin centroids for
both stripping and pick-up cases are reproduced in the following section.
\section*{II. THEORETICAL OUTLINE}
The isospin centroids, $E^{\pm}_{T}$ (superscripts +,$-$ indicate
stripping and pick-up  cases respectively) of residual nucleus states
having isospin T, obtained via single particle stripping and
pick-up reactions on a target state with isospin $T_0$ are given [6,7] by\\
\noindent
$
E^{+}_{T_>}~-~E^+ (riz) $
$$=\frac
{\sum_k\{<H^{00}_{ik}>_{Tar}+<H^{01}_{ik}>_{Tar}+(N_i-\delta_{ik})q^{+}_{T_>}(
k)
\overline{W^{T=1}_{ik}}+(N_i+\delta_{ik})r^{+}_{T_>}(k)\overline
{W^{T=0}_{ik}}\}}{<\rho_i~neutron~ holes>_{Tar}}, \eqno{(1)}$$
\noindent
$
E^{+}_{T_<}~-~E^{+}(riz) $
$$= ~\frac{\sum_k { \{ <H^{00}_{ik}>_{Tar}-(\frac
{T_0+1}{T_0})<H^{01}_{ik}>_{Tar}+(N_i-\delta_{ik})q^{+}_{T_<}(k)\overline
{W^{T=1}_{ik}}+(N_i+\delta_{ik})r^{+}_{T_<}(k)\overline
{W^{T=0}_{ik}}\} } }
{<\rho_i~proton~holes>_{Tar}+\frac{1}{2T_0}\{<\rho_i~proton~holes
>_{Tar}-<\rho_i ~neutron~holes>_{Tar}\}}, \eqno{(2)}$$
$$ E^{-}_{T_>}~-~E^{-}(riz)
=\frac{\sum_k{\{<H^{00}_{ik}>_{Tar}-<H^{01}_{ik}>_{Tar}\}}}{<\rho_i~protons>_{
Tar}},~~~~~~~~~~~~~~~~~~~~~~~~~~~~~~~~~~~~~~~~~~~~~~~ ~~~  \eqno{(3)}$$
 and\\
\noindent
$ E^{-}_{T_<}~-~E^{-}(riz) $
$$= \frac{\sum_k {\{<H^{00}_{ik}>_{Tar}+(\frac
{T_0+1}{T_0})<H^{01}_{ik}>_{Tar}\}}}{<\rho_i~neutrons>_{Tar}+\frac
{1}{2T_0}\{<\rho_i~neutrons>_{Tar}-<\rho_i~protons>_{Tar}\}}
\eqno{(4)} $$
In these equations, $T_>~\equiv ~T_0+\frac{1}{2}$ ; $T_<~\equiv
 ~T_0-\frac{1}{2}$
 ; the summation index $k$ runs over all the active orbits in the
 target state while $i$ refers to the $\rho_i(\equiv
 j\frac{1}{2})$ orbit
into (from) which the nucleon transfer occurs.  Further
$$ N_i~=~2j_i +1~~ ; \eqno{(5)}$$
$$q^{+}_{T}(k)~=~ \frac {3}{4}n_{k} + \frac
{f(T)T_{0k}}{2T_0}
 ~~ ; \eqno{(6)}$$
 $$ r^{+}_{T}(k)~=~\frac {1}{4}n_{k}- \frac
 {f(T)T_{0k}}{2T_0}~~;\eqno{(7)}$$
$n_k$ = number of nucleons in the $k$th active orbit in the
target state;
$$ f(T)~=~ T(T+1)-\frac {3}{4}-T_0(T_0+1) \equiv
\left\{ \begin{array}{ll}
T_0~for~T_> & \\
-(T_0+1)~for~T_<~~~; &
\end{array} \right.\eqno{(8)} $$
$T_{0k}$ = partial contribution of nucleons in the $k$th active
orbit towards the target state isospin;\\
$E^{\pm}(riz)~=~ E_0 \pm \epsilon_i$, with $E_0$ being the target
state energy and $\epsilon_i$, the single particle energy of
transferred nucleon with respect to the chosen inert core.\\
$\overline{W^{T=1}_{ik}}$ and $\overline {W^{T=0}_{ik}}$ in
equations (1) and (2) are $(2J+1)$-weighted averages of
two-body effective interaction matrix elements, $W^{JT}_{ikik}$, in isotriplet 
and isosinglet states,
respectively, of one nucleon in the $i$th orbit and another in
the $k$th orbit.
$$ <H^{00}_{ik}>_{Tar}=-\frac{1}{2} (1 +
\delta_{ik})E^{(2)}_{Tar}(i-k)\eqno{(9)} $$
where $E^{(2)}_{Tar}(i-k) $ is the total two-body interaction
energy of active nucleons in the $i$th orbit with those in the
$k$th orbit in the target state. $<H^{01}_{ik}>_{Tar}$ is the
isovector two-body correlation  term given by\\
\noindent
$ <H^{01}_{ik}>_{Tar}  $
$$ = \frac{1}{2} \sum_{\gamma}<Target
state\mid (2\gamma +1)^{1/2}W^{\gamma}_{ikik}[\{(A^{\rho_k }\times
A^{\rho_i})^{\gamma} \times B^{\rho_k}\}^{\rho_i} \times
B^{\rho_i}]^{01}\mid Target state>\eqno{(10)} $$
where the symbols $A^{\rho}$, $B^{\rho}$ etc. have their usual
meanings [6,7].  This term has, so far, defied an analytical
evaluation.  But as can be seen from equations (1) through (4),
this term can be eliminated by suitably combining any two of
these.\\
Usually both $T_<$ states and $T_>$ states are not fully populated in a
single particle transfer reaction, either because of experimental
limitations or because one of these is, sometimes, theoretically
untenable.  In these circumstances, it is more advisable to look
for one stripping and one pick-up reaction on the same target to
get rid of this two-body correlation term.\\
It may, however, be mentioned that for target states having
single nucleon occupancy and for those with larger nucleon
occupancy but isospin zero, the isovector two-body correlation
term, $<H^{01}>_{Tar}$, identically vanishes.
\section*{III. CALCULATIONAL PROCEDURE}
The set of experimental data which is of interest to us
is a pair of one stripping and one pick-up reaction on the same
target nucleus with stripping (pick-up) of a nucleon taking
place into (from) a specifically chosen orbit. Depending on the isospins
of residual nucleus states `seen' via these experiments, two of
the equations (1) through (4) are combined to eliminate the
isovector two-body correlation term, $<H^{01}>_{Tar}$. This
procedure essentially provides us with one equation involving $\overline
{W^{T=1}_{ik}}$ and $\overline {W^{T=0}_{ik}}$ as parameters
($i$ referring to the particular transfer orbit and $k$ running
over all the active orbits in the target state).\\
For a given transfer orbit, belonging to a major shell, usually
a large number of experiments are available in literature with
different nuclei as targets so that the number of linear
equations having the same set of parameters,
$\overline{W^{T}_{ik}}$, exceeds the number of variables. The best
values of the parameters are, therefore, found by making a
least-squares -fit. \\
As far as the coefficients of $\overline {W^{T=1}_{ik}}$ and
$\overline {W^{T=0}_{ik}}$ and the other quantities occurring in
the linear equations resulting after the elimination of
$<H^{01}>_{Tar}$  are concerned, their evaluation is
quite simple.  The isospin centroids, $E^{\pm}_{T}$, are
calculated from the experimentally measured energies and
strengths (spectroscopic factors) of various states of the
residual nucleus (having same isospin T). $E(riz)$,
$E^{(2)}_{Tar}$ and the single particle energies of nucleons in
different orbits with respect to the postulated inert core are
calculated with the help of binding energy data. The values of quantities
like $n_k$, $T_{0k}$ etc. are based on the assumption of a pure
configuration for the target state, and the denominators on the
right hand sides of the equations (1) through (4) are obtained
from the well-known non-energy-weighted sum rules [8].
\section*{IV. RESULTS AND DISCUSSION}
Although the equations (1) through (4) enable us to deal with
any target state having a multishell configuration, we have
presently restricted ourselves to target states having single
active orbit. We believe that taking up the simpler applications
as a first step would provide us with a better understanding of
the underlying principles leading to the above equations.  This
would also help us in the understanding of the effect of
enlarging the configuration space, on the two-body effective
interactions when we subsequently study the cases of
multishell target states.\\
The present work concentrates on the s-d shell region of the
periodic table. Experimental data (energies, JT values and
spectroscopic factors of residual nucleus states) were collected
from literature [9-40] for equivalent transfer (both stripping
and pick-up) of a nucleon involving the $1d_{5/2}$, $2s_{1/2}$
and $1d_{3/2}$ orbits of various nuclei as targets. Table I
gives a list of the reactions and experimental centroids used in
the present analysis.\\
As mentioned earlier, we consider only one orbit to be active
in the target state and the nucleon stripping (pick-up) taking
place into (from) this very orbit. For transfer to $1d_{5/2}$, $2s_{1/2}$ and
$1d_{3/2}$ orbits, we treat $^{16}$O, $^{28}$Si and $^{32}$S,
respectively, as the inert cores. Following the procedure outlined in the 
previous section then provides us
with three different sets of equations for the average
interaction parameters, one set for $\overline
{W^{T=0,1}_{d_{5/2}d_{5/2}}}$, second for
$\overline{W^{T=0,1}_{s_{1/2}s_{1/2}}}$ and the third for
$\overline{W^{T=0,1}_{d_{3/2}d_{3/2}}}$.
The best fitted values of the parameters obtained from these
equations are presented in Table II. The root-mean-square
deviations for the three sets of fitting done in the present
calculation are 0.673, 0.267 and 0.196 MeV respectively. For comparison, the 
values of the interaction parameters calculated from the matrix elements
obtained in some other works [41,42] are also given in the table.\\
As can be seen from table II, the presently calculated values of
average interaction parameters are in close agreement with our
previous results obtained using data from stripping reactions
alone. There is also a reasonable agreement between our results and
the average parameters obtained from the modified surface delta
interaction (MSDI) matrix elements of Halbert et al. [42]. The
average effective interaction parameters obtained by these
workers using radial integral parameterization (RIP) and those
obtained by Kuo and Brown [41] show large differences with our
results.  These differences could perhaps be partly due to larger
vector space and configuration mixing used by these workers while we work in a
truncated space limited to one active orbit only and even in
multishell target states allowing more than one active
orbit, configuration mixing is not permissible. \\
We would like to point out that the basic merit of the present sum rules
approach is its being analytical in nature. Instead of doing a mixed
configuration calculation, trying to outdo a similar calculation done
earlier, we are presenting an alternative simple approach to handle
nuclear spectroscopy of the final nucleus involved in a single particle
transfer reaction.\\
Equations, (1)-(4), of the manuscript, relating the isospin centroids to
the effective interaction parameters, both in stripping and pick-up
reactions constitute useful constraints for researchers working in the
current area.\\
\noindent
{\bf ACKNOWLEDGMENT}\\
We would like to thank the University Grants Commission (India)
for financial support to this work.
\vfill \eject
\begin{center}
Table I. List of the experiments from which the centroids have
been obtained for the present study.\\
\vskip .5truecm
\begin{tabular}{l c c c c c c c} \hline\hline
Transfer & Target & Stripping & \multicolumn{2}{c}{Centroid}
& Pick-up & \multicolumn{2}{c}{Centroid}\\ \cline{4-5}
\cline{7-8}
orbit &  & Reaction & Isospin & Value & Reaction & Isospin &
Value\\
& & [Reference]& & (MeV) & [Reference] & & (MeV)\\ \hline
$1d_{5/2}$ & $^{17}$O & $(^{3}$He,$d)$[9] & $T_<$ & 2.214 & ---
&---& ---\\
& & $(^{3}$He,$d)$[9] & $T_>$ &4.176 &---& ---& ---\\
& $^{18}$O & $(^{3}$He,$d)$[10] & $T_<$ & 2.436 & --- &--- &---\\
& & $(^{3}$He,$d)$[10] & $T_>$ & 7.630 &--- &--- &---\\
& $^{19}$F & $(d,p)$[11] & $T_>$ & 1.491 & $(d,$$^{3}$He)[12] &
$T_>$ & 3.220\\
& & $(^{3}$He,$d)$[13] & $T_>$ & 10.814 & $(d$, $^{3}$He)[12] & $T_>$
& 3.220\\
& $^{20}$Ne & $(d,p)$[14] & $T_>$ & 1.190 & --- & --- &---\\
& & $(d,n)$[15] & $T_>$ & 1.410 & --- &--- &---\\
& $^{21}$Ne & $(d,p)$[16] & $T_>$ & 3.520 &$(p,d)$[16] & $T_<$ &
2.121\\
 & $^{22}$Ne & $(d,n)$[15] & $T_>$ & 7.890 & $(d$,$^{3}$He)[12]&
 $T_>$ & 0.000\\
 & & $(d,n)$[15] & $T_<$ & 1.870 & $(d$,$^{3}$He)[12] & $T_>$ &
 0.000\\
 & $^{23}$Na & $(d,n)$[17] & $T_<$ & 2.204 & $(^{3}$He,$\alpha
 )$[18] & $T_>$ & 2.916\\
 & & $(^{3}$He,$d)$[19] & $T_>$ & 10.086& $(^{3}$He,$\alpha)$[18] &
 $T_<$ & 2.532\\
 & $^{24}$Mg & $(^{3}$He,$d)$[20] & $T_>$ & 0.308 & --- & ---& ---\\
 & $^{25}$Mg & $(^{3}$He,$d)$[21] & $T_>$ & 1.208 &
 $(d$,$^{3}$He)[22] & $T_>$ & 1.056\\
 & & $(^{3}$He,$d)$[21] & $T_<$ & 0.664 & $(d$,$^{3}$He)[22] & $T_>$
 & 1.056\\
 & $^{26}$Mg & $(^{3}$He,$d)$[23] & $T_<$ & 0.000&
 $(d$,$^{3}$He)[22] & $T_>$ & 0.011\\
 & & & & & & & \\
 $2s_{1/2}$ & $^{29}$Si & $(d,p)$[24] & $T_>$ & 1.484 & --- &---
 &---\\
 & & $(^{3}$He,$d)$[25] & $T_<$ & 0.627 & --- &---&---\\
 & & $(^{3}$He,$d)$[25] & $T_>$ & 2.314 & --- & --- &---\\
& $^{30}$Si & $(^{3}$He,$d)$[26] & $T_<$ & 0.000 & $(p,d)$[27] &
$T_<$ & 0.000\\
& $^{31}$P & $(^{3}$He,$d)$[28] & $T_<$ & 1.151 &
$(^{3}$He,$\alpha)$[29] & $T_<$ & 0.600\\
& & & & & & &\\
$1d_{3/2}$ & $^{33}$S & $(d,p)$[30] & $T_>$ & 2.480 & --- &---
&---\\
& & $(^{3}$He,$d)$[31] & $T_<$ & 0.289 & --- &--- &---\\
& $^{34}$S & $(d,p)$[32] & $T_>$ & 0.000 & $(^{3}$He,$\alpha)$[33]
& $T_<$ & 0.259\\
& & $(^{3}$He,$d)$[34] & $T_<$ & 0.000 & $(^{3}$He,$\alpha)$[33] & $T_<$
& 0.259\\
& $^{35}$Cl & $(d,p)$[35] & $T_>$ & 0.862 & $(d$,$^{3}$He)[36]&
$T_>$ & 0.000\\
 & & $(^{3}$He,$d)$[37] & $T_<$ &1.945 &$(d$,$^{3}$He)[36] &$T_>$ &
 0.000\\
& $^{36}$Ar & $(d,p)$[38] & $T_>$ & 0.000 & ---& ---& ---\\
& & $(d,n)$[39] & $T_>$ & 0.000 & --- &--- &---\\
& $^{37}$Cl & $(^{3}$He,$d)$[37] & $T_<$ & 1.915 & $(p,d)$[40] &
$T_>$ & 4.299\\ \hline\hline
\end{tabular}
\end{center}
\vfill\eject
\begin{center}
Table II. Average two-body interaction parameters (in MeV).\\
\vskip .5truecm
\begin{tabular}{l c c c c c c } \hline \hline
& & & & & &       \\
& $\overline{W^{T=1}_{d_{5/2}d_{5/2}}}$ &
$\overline{W^{T=0}_{d_{5/2}d_{5/2}}}$&
$\overline{W^{T=1}_{s_{1/2}s_{1/2}}}$&
$\overline{W^{T=0}_{s_{1/2}s_{1/2}}}$&
$\overline{W^{T=1}_{d_{3/2}d_{3/2}}}$&
$\overline{W^{T=0}_{d_{3/2}d_{3/2}}}$\\ \hline
Present calc.& -0.636  & -2.80  &-0.559 & -2.30 & -0.326 & -2.16\\
Previous calc.$^{a}$ &--- &---& -0.537 & -2.22& -0.356 & -2.22\\
Kuo and Brown$^{b}$ & -0.398 &-2.31 &-2.21 & -3.54 &-0.127 &
-1.68\\
Hamiltonian `RIP'$^{c}$ & -0.171 &-2.32 &-1.72 & -4.32& -0.035
&-2.38\\
Hamiltonian `MSDI'$^{c}$& -0.248 &-3.50 & -0.584 & -3.28 &
-0.266 & -3.43\\ \hline\hline
\end{tabular}
\end{center}
\vskip .5truecm
$^{a}$Reference [5].\\
$^{b}$Reference [41]. \\
$^{c}$Reference [42].

\vfill \eject

\end{document}